\documentclass[12pt]{iopart}
\usepackage{iopams}

\usepackage[dvipdfmx]{graphicx}
\begin{document}

\title[Biased diffusion on Japanese inter-firm  trading network]{
Biased diffusion on Japanese inter-firm  trading network: Estimation of sales from network structure}

\author{{Hayafumi Watanabe}$^1$, {Hideki Takayasu}$^2$ and {Misako Takayasu}$^1$}

\address{ $^1$Department of Computational Intelligence and Systems Science, Interdisciplinary Graduate School of Science and Engineering, Tokyo Institute of Technology, 
4529 Nagatsuta-cho, Midori-ku, Yokohama 226-8502, Japan \\
 $^2$ Sony Computer Science Laboratories, 3-14-13 Higashi-Gotanda, Shinagawa-ku, Tokyo 141-0022, Japan}
\ead{h-watanabe@smp.dis.titech.ac.jp }

\begin{abstract}
To investigate the actual phenomena of transport on a complex network, we
analysed empirical data for an inter-firm trading network, which consists of 
about one million Japanese firms and the sales of these firms (a sale
corresponds to the total in-flow into a node). First, we analysed the
relationships between sales and sales of nearest neighbourhoods from which we obtain a simple linear relationship between sales and
the weighted sum of sales of nearest neighbourhoods (i.e., customers). In
addition, we introduce a simple money transport model that is coherent with 
this empirical observation. In this model, a firm (i.e., customer) distributes money to its
out-edges (suppliers) proportionally to the in-degree of destinations. From intensive
numerical simulations, we find that the steady flows derived from
these models can approximately reproduce the distribution of sales of actual firms.
The sales of individual firms deduced from the money-transport model are shown to be proportional, on an average, to the real sales. \par
\end{abstract}
\pacs{89.75.Hc, 89.75.Da, 05.10.Gg}

\maketitle
\section{Introduction}
The circulation of money is often likened to the circulation of blood. 
For instance, in the middle of the 18th century, French physiocrat Fran\c{c}ois Quesnay 
introduced his economic theory \lq tableau \'{e}conomique\rq, which is one of the important foundations of modern economics, from the theory of blood circulation described by William Harvey in the 17th century \cite{blood}.
If this analogy is acceptable, what are the differences between the flow of money (i.e., the flow within society) and the flow of blood (i.e., the flow within the body) in terms of transport phenomena? \par
Transport processes such as diffusion, advection and radiation, play a fundamental role in physics and theories of transport processes are widely applied in chemistry, biology, engineering, etc.  
One transport problem that has long attracted interest is transport in complex systems such as  biological or the social systems. 
Because of recent accumulations of data regarding complex systems and developments in the theory of complex networks, we can now provide a new perspective on such problems. 
 \par 
Complex networks have been studied intensively over the past decade 
\cite{Brabashreview}-\cite{Breview}. 
These studies revealed that complex networks can be observed in a wide range of real systems both in natural and man-made.
In particular, transport phenomena in complex networks have been investigated. 
For example, random walks on complex networks have been studied from various viewpoints 
\cite{pagerank,random_complex,page_average}. 
The PageRank, which is one of the most successful indices evaluating the importance of web pages and which is applied by Internet search engines, corresponds to the steady-state density of transport caused by random walks on the World Wide Web.
Other types of transportation on complex networks have also been studied 
\cite{Trasper_phenomeona_complex_network,denki,selfavoid}. 
\par
What properties characterize the actual transport on a complex network?
A great majority of studies of transportation on complex networks has been based on theoretical approaches; however, a few studies have been involved actual transportation on complex networks. 
For example, R. Guimera \etal. revealed the nonlinear relationships between degrees of airports traffics on the world-wide-airport network \cite{hikouki},
P. Sen \etal. studied the number of trains on the Indian rail way network \cite{IndianRailWay} and A. Chmiel \etal.  investigate  on visitors on portal sites and self-attracting walks 
can well describe its properties \cite{browsing}.  
\par 
To investigate real phenomena of transport on a complex network, we focus on the transport of money on the trading network where firms correspond to network nodes and the trade relations between firms correspond to network edges. 
This system is useful for studying the real transport phenomena on the complex network, because we can estimate the total in-flow of each node by from the sales data. 
 \par
In this study, we show that we can consistently estimate the sales from the structure of the inter-firm trading network, which is analogous to estimating blood flows from the vascular structure.  In sec. II, we start by analysing the trading network data, from about 900,000 Japanese firms and  their corresponding sales data.
Next, we introduce two transport models and discuss their properties in Sec. III. 
In Sec. IV, we compare the steady-state sales of the model with the actual sales and show that flows generated by the models can reproduce the well-known Zipf's law; namely, that the cumulative distribution of sales obeys a power-law distribution with the exponent close to -1. 
Finally, we conclude with a discussion in Sec. V. 
\section{Data analysis}
 The data set was provided by Tokyo Shoko Research, Ltd. (TSR). 
 and contains about one million firms practically covering all active firms in Japan. 
For each firm, the data set contains the annual sales and a list of business partners, categorized into suppliers and customers \cite{oonishi,Aoyama}. 
\par
\begin{figure}[t]
\begin{minipage}{0.3333\hsize}
\centering
\includegraphics[width=6cm]{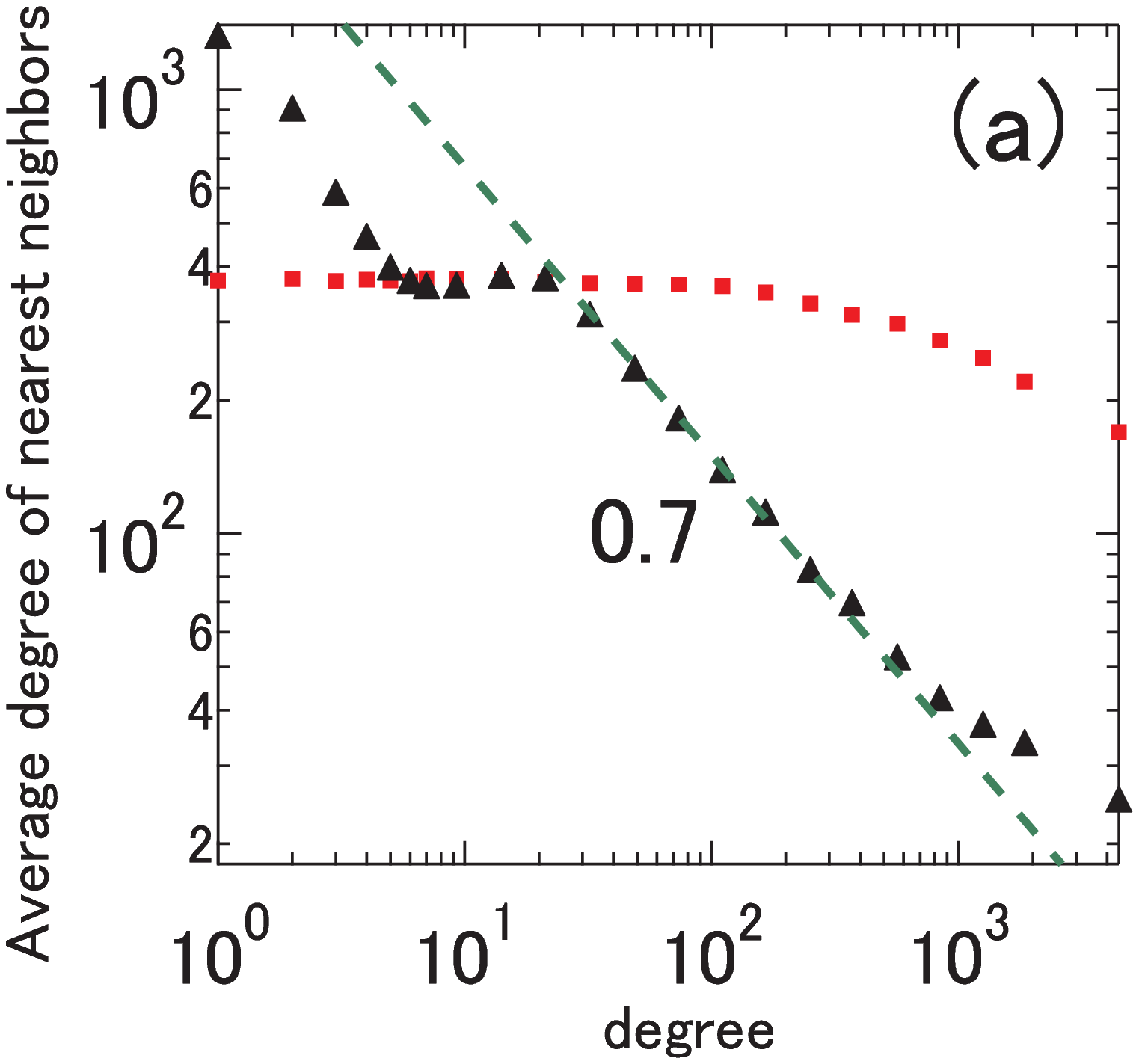}
\end{minipage}
\begin{minipage}{0.333333333\hsize}
\centering
\includegraphics[width=5.7cm]{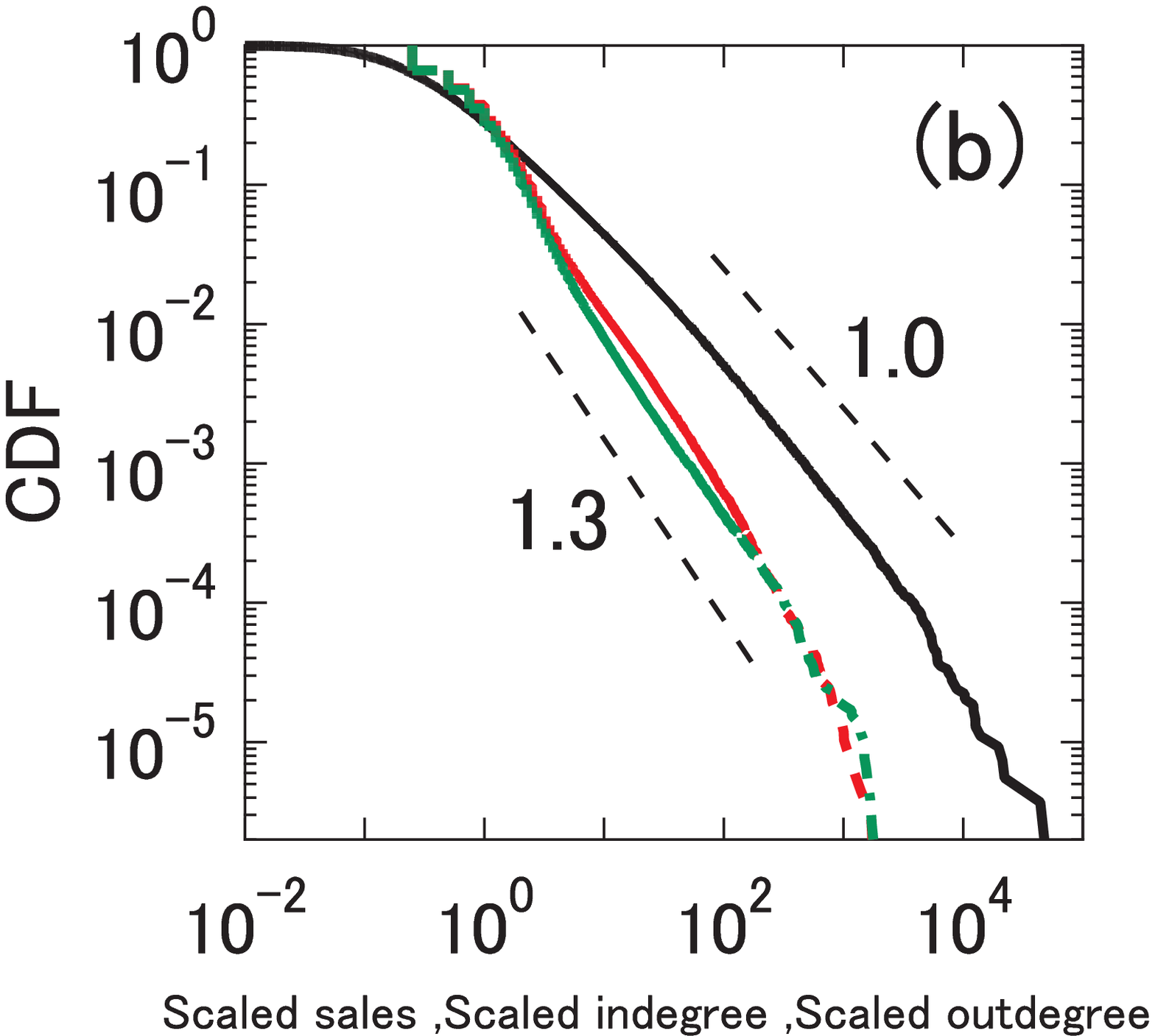}
\end{minipage}
\begin{minipage}{0.33333\hsize}
\centering
\includegraphics[width=5.8cm]{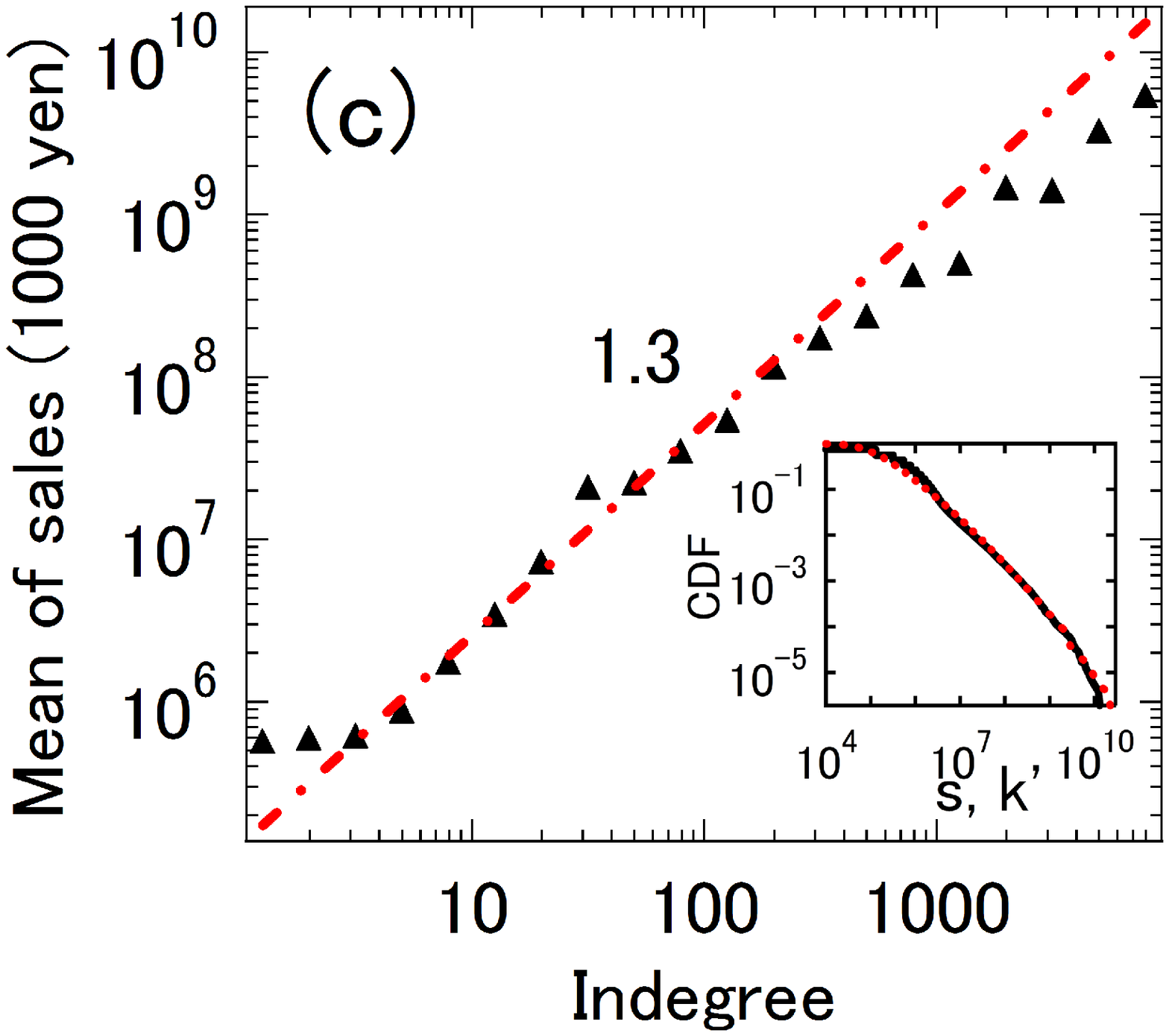}
\end{minipage}
\caption{
(a)Average degree of nearest neighbours $k_{nn}(k)$ for the LSCC of the firm network (black triangles) and that for the shuffled network(red squares). 
The LSCC of the firm network has a stronger negative degree-degree correlation than the shuffled network. \\
(b)The Cumulative distribution of scaled sales (black solid line), scaled in-degrees(red broken line) and scaled out-degrees(green dash-dotted line). Each value is scaled(i.e., divided) by the inter quartile range.
The exponent for sales is different from that for in-degrees  and out-degrees.\\
(c)The conditional mean of sales given in-degree (black triangles). The red dash-dot line is $s=10^{5.1} \cdot {k^{(in)}}^{1.3}$. 
The conditional mean is proportional to the $1.3$ power of in-degrees. 
The bottom right figure shows CDFs of $s$ (black solid line) and $k'=10^{5.1} \cdot {k^{(in)}}^{1.3}$ (red broken line). These CDFs are in good agreement.\\
}
\label{basic}
\end{figure}

From this list, we generated a network (firm network), 
whose nodes are the firms and the edges are defined by the following rule: If the i-th firm buys something from the j-th firm, or equivalently if money flows the i-th to the j-th, we connect from the i-th to the j-th with a directed link\cite{oonishi}.  
Figure \ref{basic}(a) shows the average degree of the nearest neighbours, which is denoted by a function of degree k: $k_{nn}(k)$ \cite{knn}.
The data shown in this figure confirms that the firm network has a negative degree-degree correlation. \par
To clarify the properties of this firm network, we perform a parallel analysis using an artificial random network having the same degree distribution as the firm network \cite{oonishi,chain,chain2}.
We generate an artificial random network by using the Markov-chain Monte Carlo switching algorithm\cite{shuffle} which is used to repeatedly choose the edge pairs randomly  and switch from \lq $X1 \to Y1$ and $X2 \to Y2$\rq \quad to  \lq $X1 \to Y2$ and $X2 \to Y1$\rq \quad until the network is well randomized. 
This artificial network, which we call the shuffled network in this paper, is an almost uncorrelated network having the same degree distribution as the original firm network. 
The red line in figure \ref{basic}(a) shows the degree-degree correlation of this network. 
By comparing the differences in behaviour between the real firm network and the shuffled network, we can check the effect of a correlation.  

\subsection{Statistical properties of individual firms}
We begin by investigating the properties of individual firms.
figure \ref{basic} (b) shows cumulative distributions $s$, $k^{(in)}$ and $k^{(out)}$, where $s$ is the annual sales in 2005, $k^{(in)}$ is the in-degree and $k^{(out)}$ is the out-degree. 
The data shown in the figure indicates that $s$, $k^{(in)}$ and $k^{(out)}$ obey the following power-law cumulative distribution functions (CDFs):
\begin{eqnarray}
P(>s) \propto s^{-\alpha_s} \label{cdf1} \\
P(>k^{(in)}) \propto \left(k^{(in)}\right)^{-\alpha_{in}} \label{cdf2} \\
P(>k^{(out)}) \propto \left(k^{(out)} \right)^{-\alpha_{out}} \label{cdf3} 
\end{eqnarray}
with exponents $\alpha_s=1.0$ ,  $\alpha_{in}=1.3$ and  $\alpha_{out}=1.3$. For sales $s$, this empirical fact, which is well known as Zipf's law, is observed in various countries
\cite{parete,usa,Fujiwara,oonishi}. \par
Next, we investigate the correlation between sales and degrees. 
We calculate the conditional mean of $s$ as a function of $k^{(in)}$.
As shown in figure \ref{basic} (c), for a large in-degree $k^{(in)}$, $<s>_{k^{(in)}}$ can be described as a power law 
\begin{center}
\begin{equation}
<s>_{k^{(in)}} \propto {k^{(in)}}^{\beta_{s|k}} \label{real_indegree_sales}
\end{equation}
\end{center}
with $\beta_{s|k} \approx 1.3$, where $<s>_{k^{(in)}}$ is the conditional mean of $s$ for given $k^{(in)}$.
These results imply that the mean of 'sales per in-degree' increases with increasing in-degree. \par
Roughly speaking, we can theoretically derive a relationship among $\alpha_{in}$, $\alpha_s$, $\beta_{s|k}$ 
as a result of transformation of random variables.
Assuming that, $k^{(in)}$ obeys the power-law distribution with the probability density function (PDF)
\begin{equation}
p_{k^{(in)}}(k^{(in)}) \propto {k^{(in)}}^{-\alpha_{in}-1}
\end{equation},
and $s$ and $k^{(in)}$ satisfy the power law relationship
\begin{equation}
s \propto {k^{(in)}}^{\beta_{s|k}},
\end{equation}
then by changing the variables, the PDF of $s$ becomes
\begin{equation}
p_{s}(s) \propto p_{k^{(in)}}(k^{(in)})\cdot \left| \frac{dk^{(in)}}{ds} \right| \propto s^{1/\beta_{s|k}(-\alpha_{in}-1)} \cdot s^{1/\beta_{s|k}-1} \propto s^{-\alpha_{in}/\beta_{s|k}-1}.
\end{equation}
Thus, we get the following nontrivial relationship between power law indices:
\begin{equation}
\beta_{s|k}=\alpha_{in}/\alpha_{s} \label{sisuu}.
\end{equation}
This relationship is consistent with the observed values.
\subsection{Nearest-neighbourhood correlations}
 We now consider the relationships between the sales of a firm and the sales of its customers. Customers of m-th node are defined by the nodes whose out-going edges reach the i-th node. 
  We introduce two kinds of weighted sums of customer sales, $s^{(1)}_{m}$ and $s^{(2)}_{m}$:
\begin{eqnarray}
s^{(1)}_{m} \equiv \sum^{N}_{i=1}A_{im}\frac{1}{\sum^{N}_{j=1}Aij}s_i=\sum^{N}_{i=1}A_{im}\frac{s_i}{{k_{i}}^{(out)}} \label{ww1} \\
s^{(2)}_{m} \equiv \sum^{N}_{i=1}A_{im}\frac{\sum^{N}_{l=1}A_{km}}{\sum^{N}_{j=1}A_{ij}\sum^{N}_{l=1}A_{lj}}s_i=\sum^{N}_{i=1}A_{im}\frac{k^{(in)}_{m}}{\sum^{N}_{j=1}A_{ij}k^{(in)}_{j}}s_i   \label{ww2}
\end{eqnarray}
where, A is an $N \times N$ adjacency matrix defined by 
\begin{eqnarray}
A_{ij}=
\cases{
 1 & if there is an edge from i to j\\
 0 & otherwise \\}
\end{eqnarray}
\par
In the case of $s^{(1)}$, a customer distributes money among all its suppliers evenly.
However, in the case of $s^{(2)}$, a customer distributes money among its suppliers in proportion to suppliers' in-degree.
We calculate conditional mean of $s^{(1)}$ given $s$, denoted by $<s^{(1)}>_s$, and conditional mean of $s^{(2)}$ given $s$, denoted by $<s^{(2)}>_s$, functions of $s$:
\begin{eqnarray}
<s^{(1)}>_{s}&=&\frac{\sum_{l \in \left\{l|s-\delta_s^{(b)} \leq s_l < s+\delta_s^{(u)}\right\}}s_l^{(1)}}{\sum_{l \in \left\{l|s-\delta_s^{(b)} \leq s_l < s+\delta_s^{(u)}\right\}}1} \nonumber \\ 
 &=&\frac{\sum_{l \in \left\{l|s-\delta_s^{(b)} \leq s_l < s+\delta_s^{(u)}\right\}}\left\{\sum^{N}_{i=1}A_{il}\frac{s_i}{{k_{i}}^{(out)}} \right \}}{\sum_{l \in \left\{l|s-\delta_s^{(b)} \leq s_l < s+\delta_s^{(u)}\right\}}1}, \label{w1} \\
<s^{(2)}>_{s}&=&\frac{\sum_{l \in \left\{l|s-\delta_s^{(b)} \leq s_l < s+\delta_s^{(u)}\right\}}s_l^{(2)}}{\sum_{l \in \left\{l|s-\delta_s^{(b)} \leq s_l < s+\delta_s^{(u)}\right\}}1} \nonumber \\
&=&\frac{\sum_{l \in \left\{l| s-\delta_s^{(b)} \leq s_l < s+\delta_s^{(u)}\right\} \left\{\sum^{N}_{i=1}A_{il}\frac{k^{(in)}_{l}}{\sum^{N}_{j=1}A_{ij}k^{(in)}_{j}}s_i\right\}}}{\sum_{l \in \left\{l|s-\delta_s^{(b)} \leq s_l < s+\delta_s^{(u)}\right\}}1},  \label{w2}  
 \end{eqnarray}
where, $\delta_s^{(b)}, \delta_s^{(u)} << s$. 
In this paper, we estimate other conditional means 
$<\cdot>_\cdot$ 
in a similar manner. \par
\begin{figure}[t]
\begin{minipage}{0.3333\hsize}
\centering
\includegraphics[width=6cm]{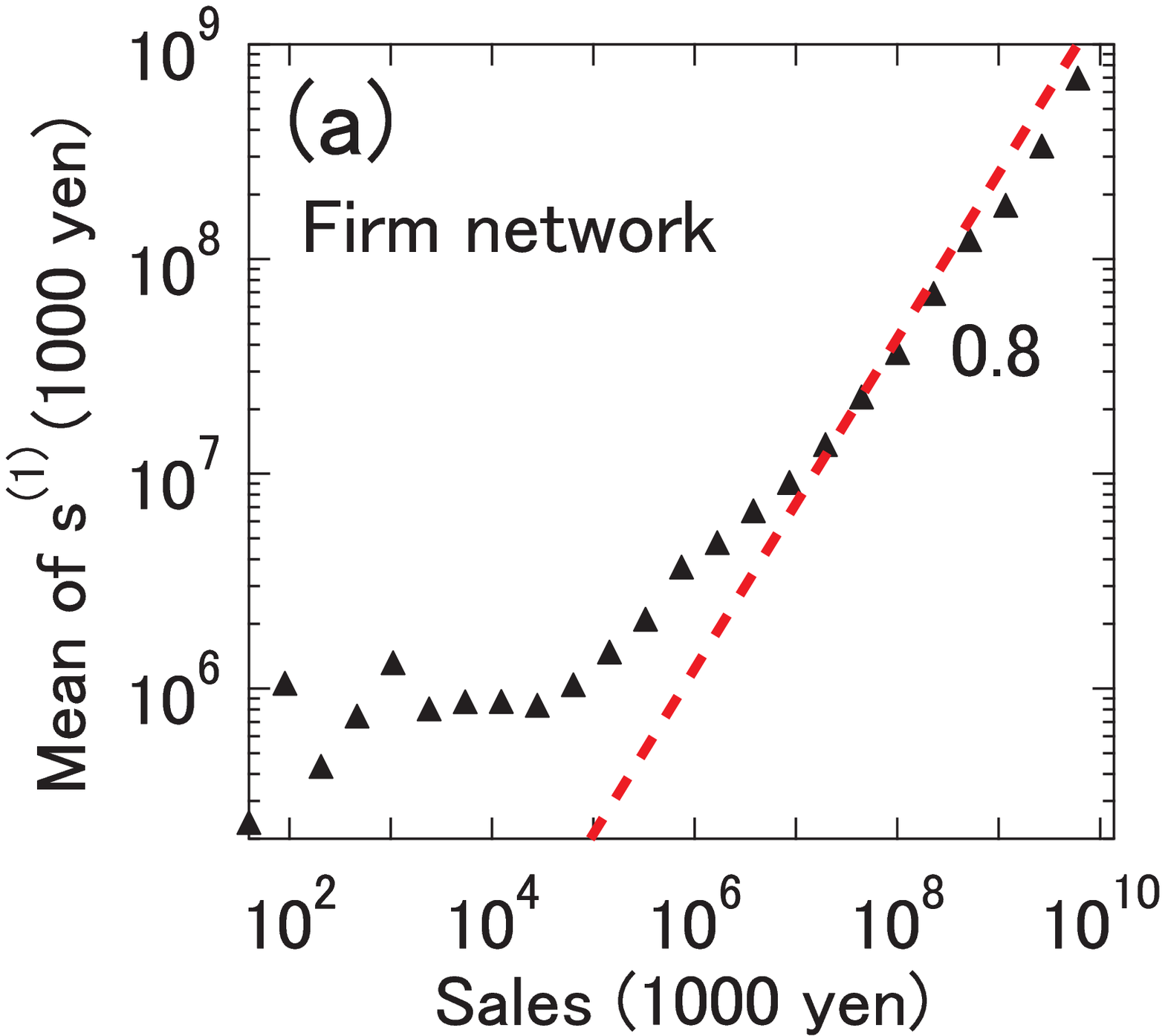}
\end{minipage}
\begin{minipage}{0.333333\hsize}
\centering
\includegraphics[width=6cm]{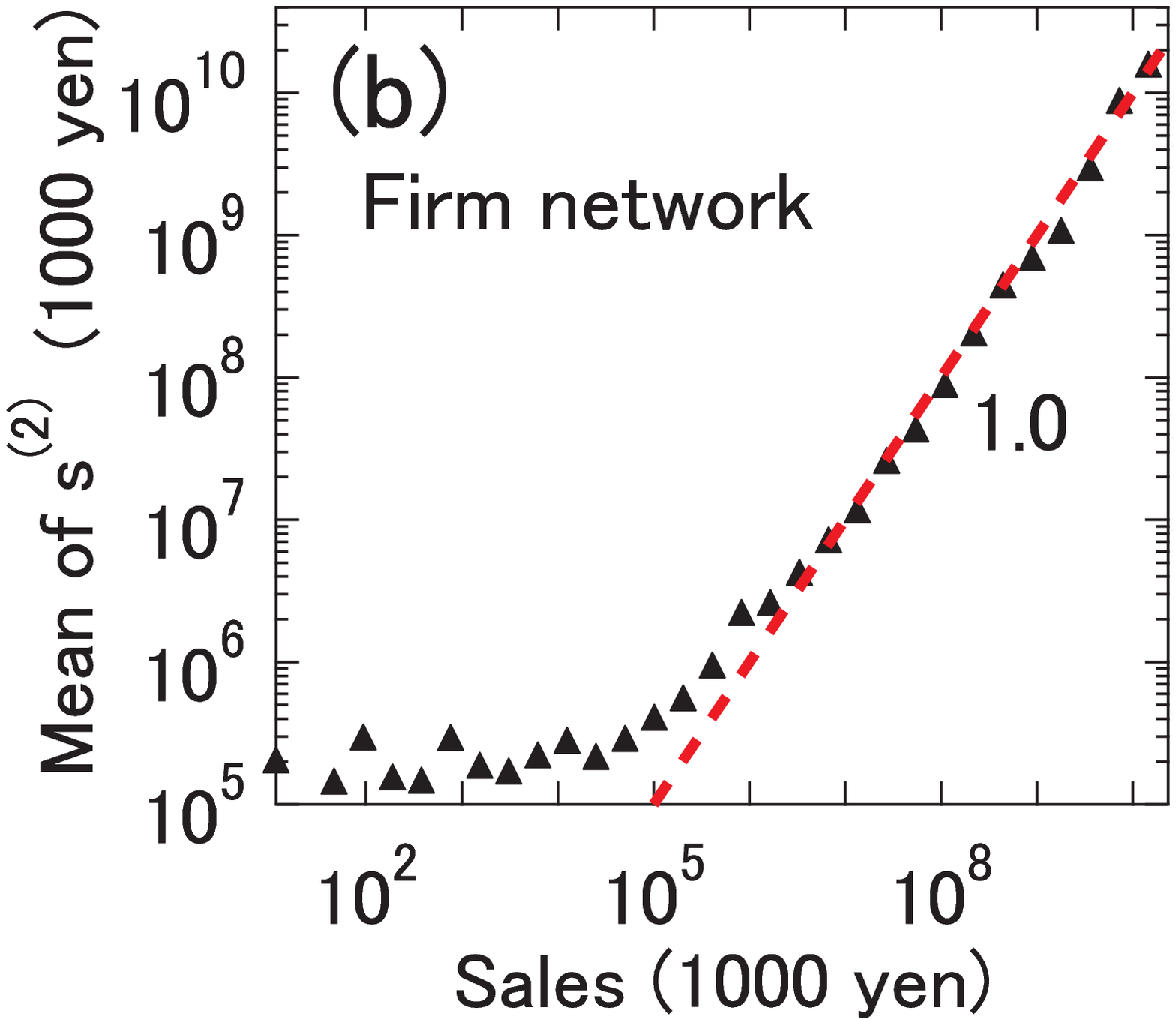}
\end{minipage}
\begin{minipage}{0.33333\hsize}
\centering
\includegraphics[width=6cm]{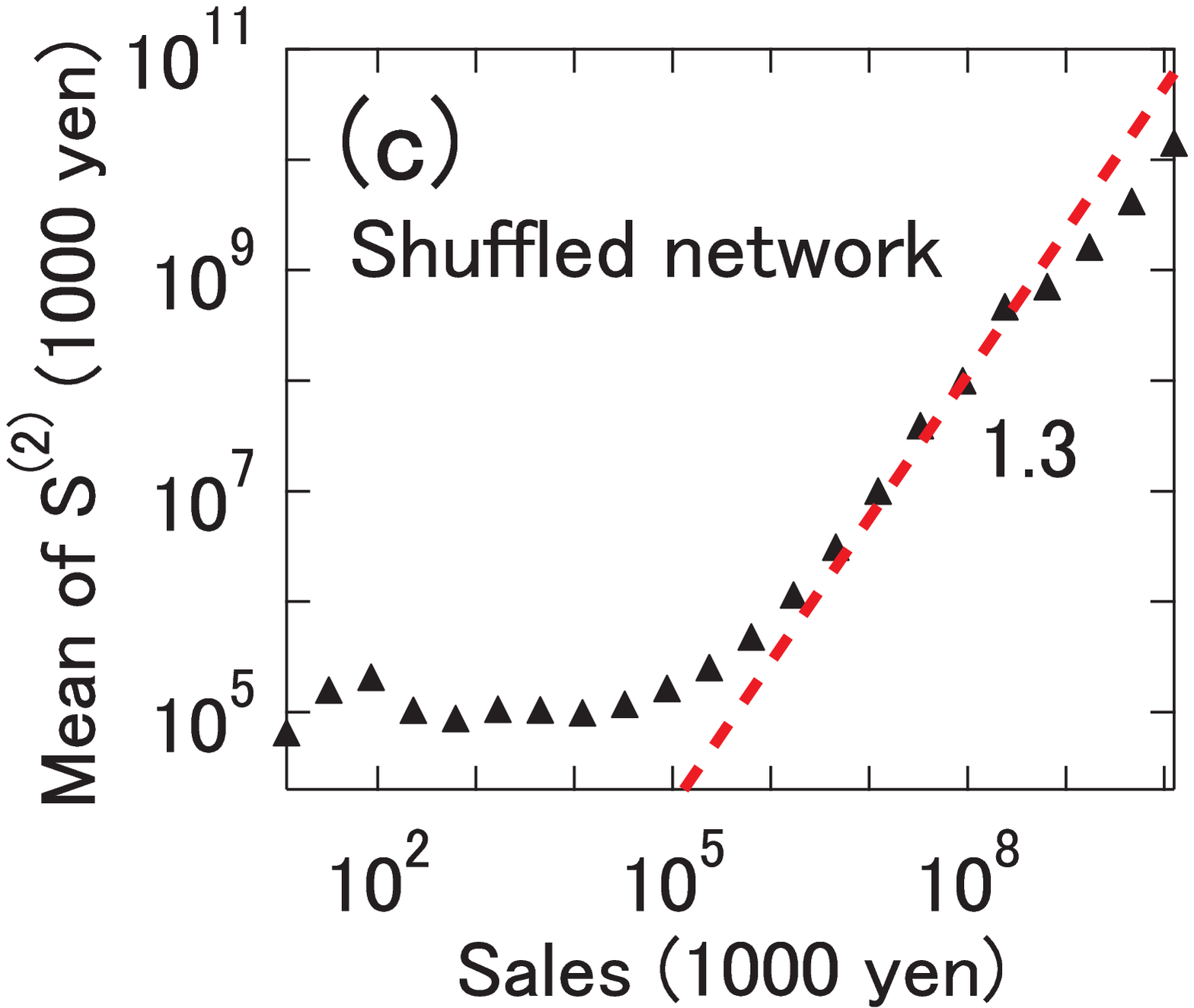}
\end{minipage}
\caption{
(a)-(c) Correlations between sales and $s^{(i)} \quad (i=1,2)$.  
(a) $<s^{(1)}>_{s}$, which is the conditional mean of $s^{(1)}$ given by the sales $s$, as a function of the sales $s$ for the firm network.  
Red broken line shows $s^{(1)}=30 \cdot s^{0.8}$.
(b)  $<s^{(2)}>_{s}$, which is the conditional mean of $s^{(2)}$ given by the sales $s$, as a function of the sales $s$  for the firm network. 
Red broken line shows $s^{(2)}=s$. 
(c) $<s^{(2)}>_{s}$, which is the conditional mean of $s^{(3)}$ given by the sales $s$, as a function of the sales $s$  for the shuffled network. 
Red broken line shows $s^{(2)}=0.0045 \cdot s^{1.3}$. 
We confirm that for the firm network,  $<s^{(2)}>_s$ is almost equal to $s$ for the large sales region. In contrast, $<s^{(1)}>_s$ for the firm network and   $<s^{(2)}>_s$ for the shuffled network are not proportional to $s$.  \\
}
\label{scaling}
\end{figure}

 As shown in figures \ref{scaling}(a) and \ref{scaling}(b), for large sales $s$, $<s^{(1)}>_s$ and $<s^{(2)}>_s$  can be described as power laws:  
\begin{eqnarray}
<s^{(1)}>_{s} \propto s^{0.8} \label{s1} \\
<s^{(2)}>_{s} \propto s^{1.0} \label{s2} \label{s2}. 
\end{eqnarray}
For $s^{(2)}$, in particular, we observe a simple linear relationship in the region above $10^9$ yen. Its proportionality constant is equals to about 1.  \par 
The results shown in figure \ref{scaling} (c) confirm that $<s^{(2)}>_s$ is not proportional to $s$  for the shuffled network. In other words, equation (\ref{s2}) does not hold for the shuffled network, which has the same degree distribution as the real firm network. 

\section{Simulations of transport}
\subsection{Models}
In correspondence with the local relationships given in equations (\ref{ww1}) and (\ref{ww2}), we introduce the following two models of the time evolution of locally conserved scalar quantities, $x_m(t)$. 
\\
Model-1 (PageRank model)
\begin{equation}
x^{}_{m}(t+1)=\sum^{N}_{i=1}A_{im}\frac{1}{k_i^{(out)}}x^{}_{i}(t), \label{model1}
\end{equation}
Model-2 (Biased distribution model)
\begin{equation}
x^{}_{m}(t+1)=\sum^{N}_{i=1}A_{im}\frac{k_m^{(in)}}{\sum^{N}_{j=1}A_{ij}k_{j}^{(in)}}x^{}_{i}(t). \label{model2}\end{equation} 
Note that for $k_i^{(out)}=0$ in equation (\ref{model1}) or $\sum^{N}_{j=1}A_{ij}k_{j}^{(in)}=0$ in equation (\ref{model2}), we omit the contributions of the i-th node.
\par
In equation (\ref{model1}), a node(customer) evenly distributes its scalar(money) among its outgoing edges (suppliers); i.e., Model-1 corresponds to the PageRank model \cite{pagerank}. 
However, Model-2 corresponds to a kind of biased random-walk model \cite{bias}. In equation (\ref{model2}), a node(customer) distributes its scalar(money) to its outgoing edges in proportion to the node in-degrees indicated by the outgoing edges (suppliers). 
\par
We apply these time-evolution models to two types of networks. 
The first network is the largest strongly connected component(LSCC) of the real firm network. 
A strongly connected component(SCC) is defined as the maximal subset of edges in a network such that each node can reach all others and is itself reachable from all others along a directed path\cite{scc}.
The LSCC of the firm network is defined as the SCC having the largest number of nodes when we decompose the firm network into SCCs.
 The LSCC of the firm network is the core of the firm network, containing 462,602 nodes and 2,583,620 edges. 
In addition, the previously mentioned statistical properties of the firm network holds true for the LSCC of the firm network. Note that for the both Model-1 and Model-2, 
there is no need to  consider  a boundary-condition or end effects, because of an SCC does not have exits(nodes having no out-edges) or entrances(nodes having no in-edges). 
Therefore, we can focus on the properties of models for the bulk of the network, which is why we do not use the original firm network but the LSCC of the original firm network for the first step of the numerical simulations. \par
The second network is the shuffled LSCC of the firm network, generated by the above-mentioned Markov-chain Monte Carlo switching algorithm\cite{shuffle}.
The shuffled LSCC of the firm network is an almost uncorrelated network with the same degree distribution as the LSCC of the firm network.
For our simulation, we used the shuffled LSCC of a network that consisted of a single SCC and we also assumed that the network consisted of a single SCC for the theoretical analysis.  
\par
\subsection{Properties of models}
For a strongly connected network, we consider the existence of a steady state for the Model-1 and Model-2.
Because SCCs have no outlets, the total scalar $\sum_{i=1}^{N}x^{}_i(t)=\sum_{i=1}^{N}x^{}_i(0)$
  is conserved in both models. 
For the steady state with normalization, $\sum^{N}_{i=1}x_i(0)=\sum^{N}_{i=1}x_i(t)=1$,  
both models are Markov-chains, where the probability of existence of the m-th node, $p_m(t)$ is given by $x_m(t)$ and the transition probability $Q_{mi}$ from the i-th node to the m-th node is 
$A_{im} \cdot 1/k_i^{(out)}$ and $A_{im}\cdot k_m^{(in)}/\sum^{N}_{j=1}(A_{ij}k_j^{(in)})$
for the Model-1 and Model-2, respectively. 
These Markov-chains are irreducible; i.e., there is a non-zero transition probability from any state to any other state.
This property arises because a path exists between any two nodes in the graph on the strongly connected network and the transition probability from the i-th node to the j-th node is a non-zero for the node pair the i-th and the j-th such that $A_{ij}=1$. 
In general, it is known that an irreducible Markov-chain with the finite number of states have unique steady state \cite{MC}, therefore, both models have the unique steady state for the strongly connected network.
We denote this steady state for the given strongly connected network by $p^{(s)}$. 
According to linearity, the steady state $x^{(s)}$ is obtained as 
\begin{equation}
{x_{m}^{(s)}}=\left(\sum_{i=1}^{N}x_i(0)\right) \cdot p_m^{(s)} \quad (m=1,2,\cdots,N).
\end{equation}
where $\sum^{i=1}_{N}x_i(0)$ is the total sum of the initial values. 
 \par 
\subsubsection{Model-1 (PageRank model)}
To understand the properties of our model for the LSCC of the firm network and the shuffled LSCC of the firm network, we simulate the Model-1 by equation (\ref{model1}).
Starting with the initial condition  $x_i(0)=1 \quad (i=1,2,\cdots,N)$, the CDF of $x$ converges to the steady state distribution as shown in figure \ref{cdf_page}(a).
For both cases the real and shuffled network, the distribution of 
 $x$ follows a power law with an exponent of about 1.3, which is the power-law exponent for in-degree $\alpha_{in}$. 
Thus, Model-1 is not consistent with the empirical sales distribution because the empirical sales distribution follows the Zipf's law with exponent -1.
Figure $\ref{cdf_page}$(b) shows the conditional mean of $x$ for given $k^{(in)}$ as a function of $k^{(in)}$, which is denoted by $<x>_{k^{(in)}}$.
For both networks, we get the following liner function:
\begin{equation}
<x^{}>_{k^{(in)}} \propto k^{(in)}. \label{page_s_k}
\end{equation}
This result can be explained by the mean-field solution of PageRank \cite{page_average}. 
However, it is inadequate to regard $x$ as sales because equation (\ref{page_s_k}) disagrees with the empirical relationship, equation (\ref{real_indegree_sales}). 
\par
\subsubsection{Model-2 (Biased diffusion model)}
\begin{figure}[p]
\begin{minipage}{0.5\hsize}
\centering
\includegraphics[width=8cm]{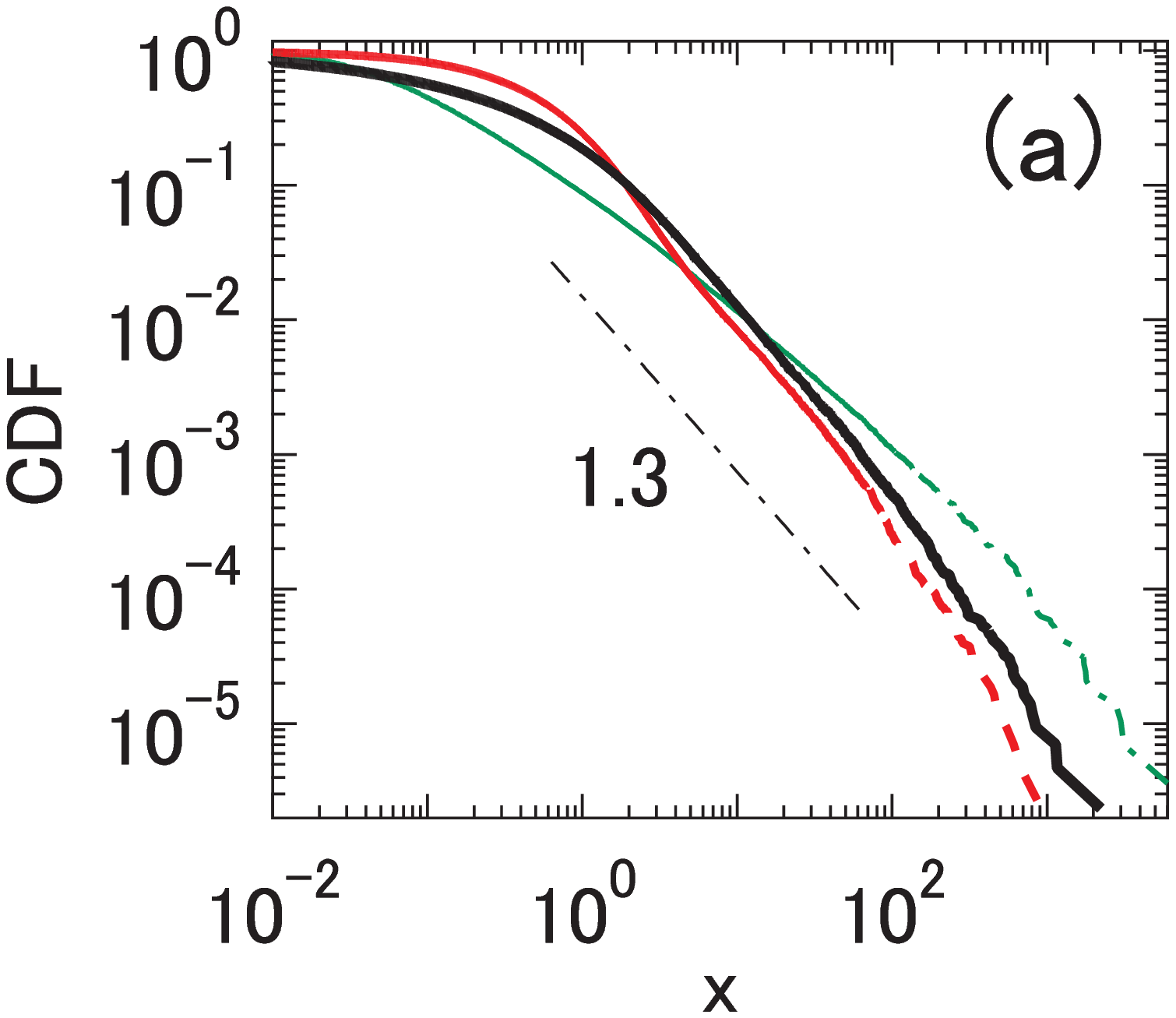}
\end{minipage}
\begin{minipage}{0.5\hsize}
\centering
\includegraphics[width=8cm]{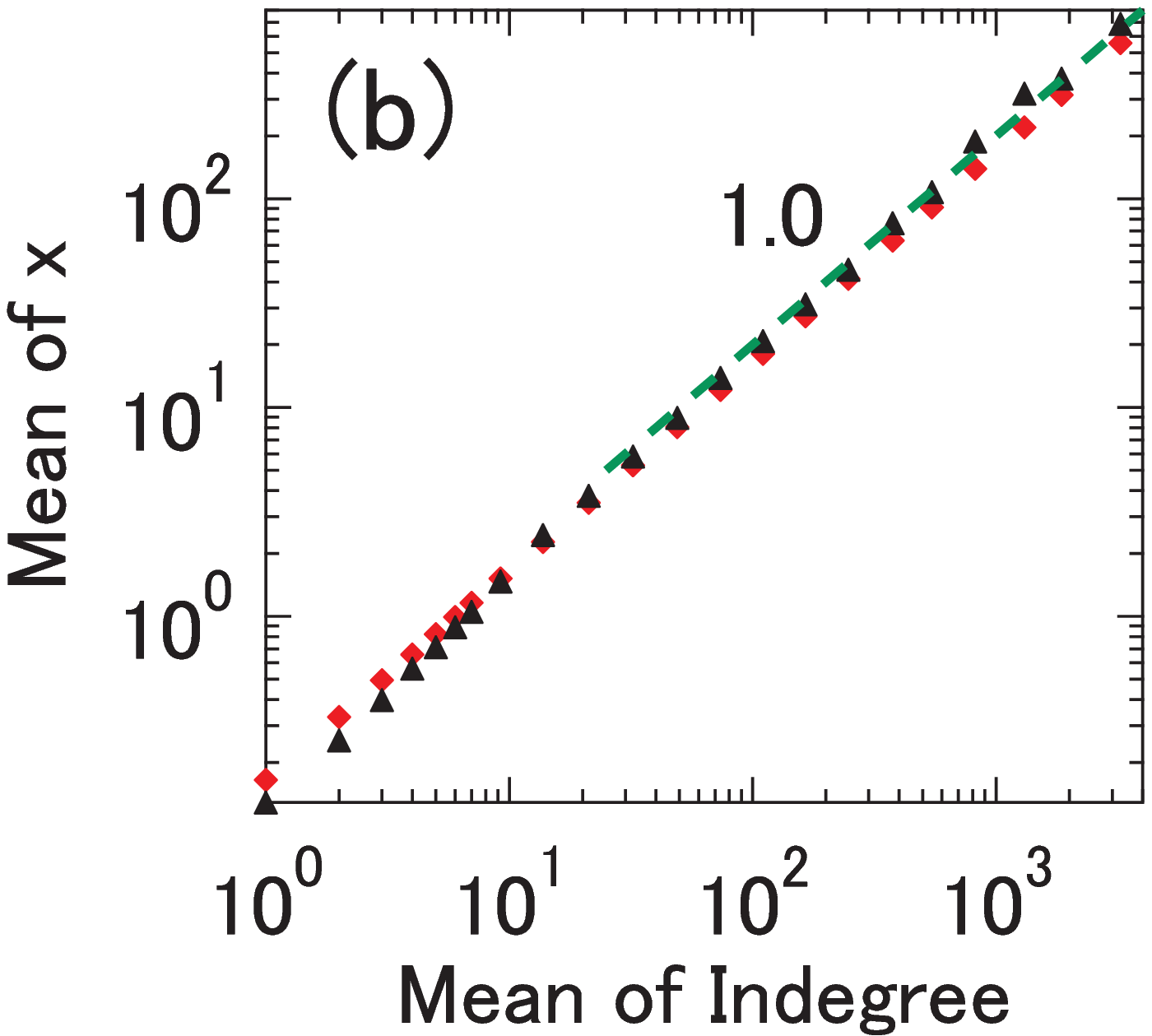}
\end{minipage}
\begin{minipage}{0.5\hsize}
\centering
\includegraphics[width=8cm]{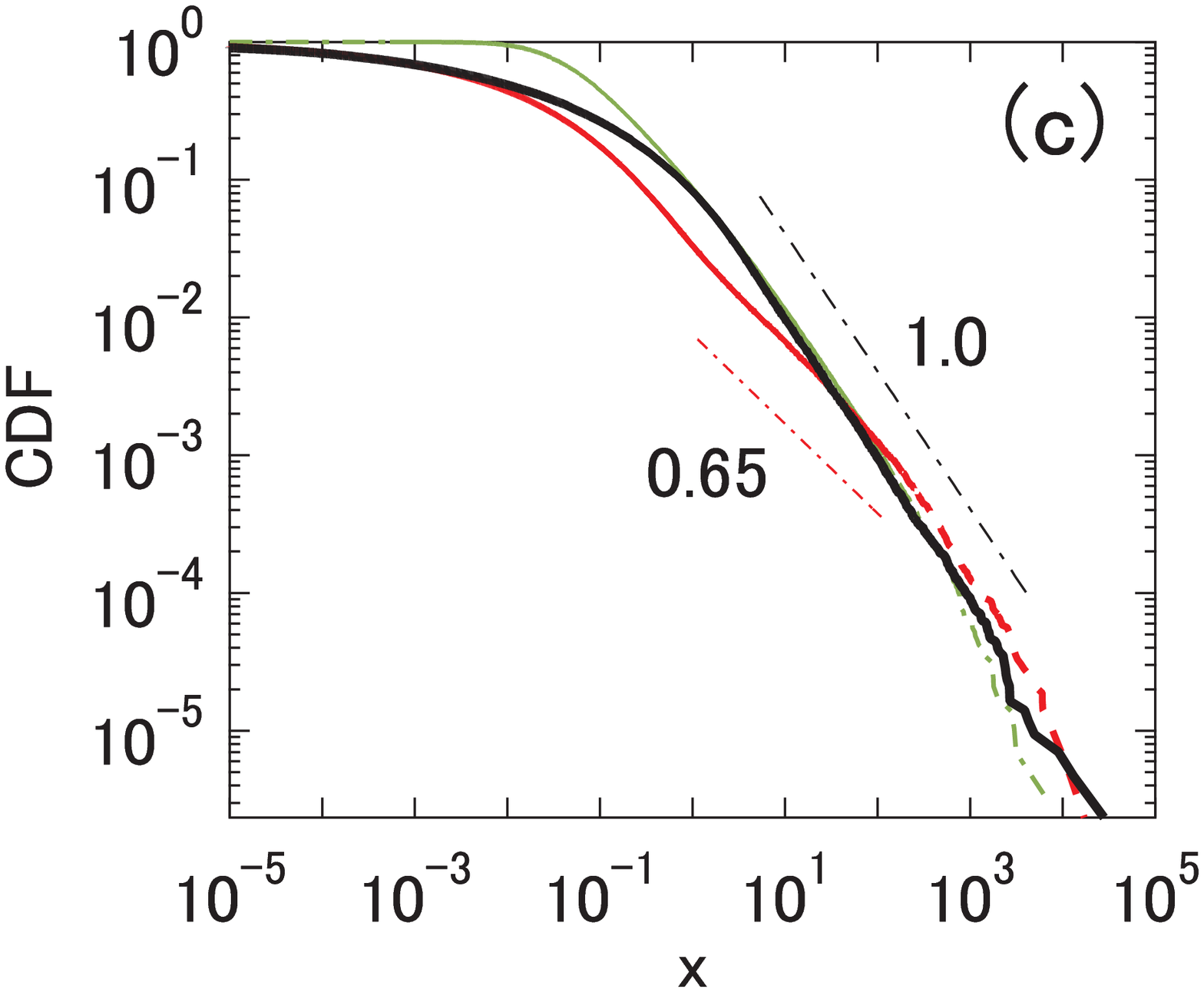}
\end{minipage}
\begin{minipage}{0.5\hsize}
\centering
\includegraphics[width=8cm]{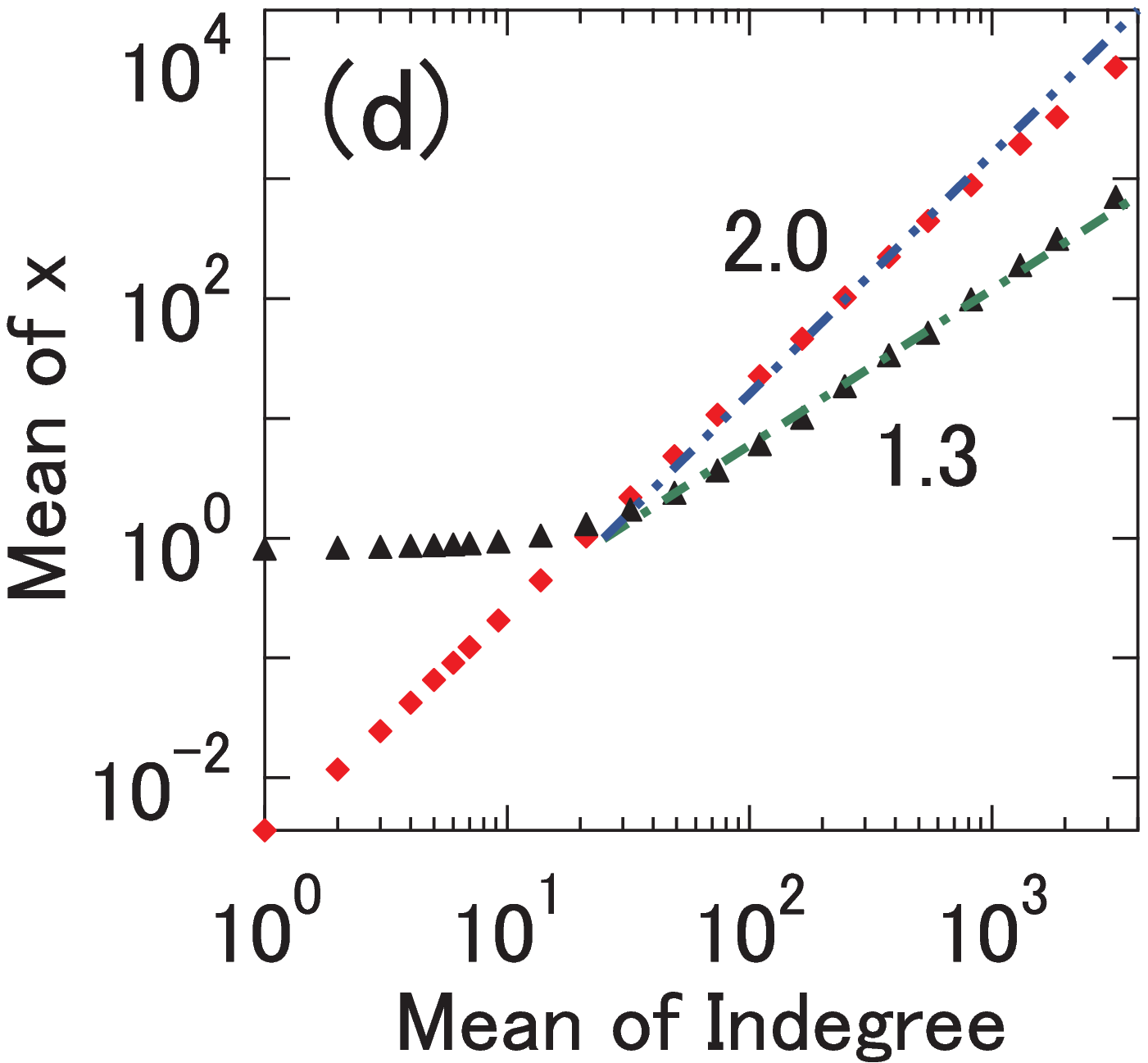}
\end{minipage}
 
\caption{
(a)CDFs of $x$ in the case of Model-1 for the large time $t$ ($t=2^{15}$). 
Shown are for the LSCC of the firm network (black solid line) and for the shuffled LSCC of the firm network (red dashed line). Sales of real firms scaled by the sum of sales are shown the green dash-dotted line.  For both networks, $x$ obeys a power law distribution with an exponent about 1.3. \\
(b)The conditional mean of $x$ given the in-degree $k^{(in)}$, denoted by $<x>_{k^{(in)}}$ in the case of Model-1.
Data shown are for the LSCC of firm network(black triangle), for the shuffled LSCC of the firm network(red diamond) and $x \propto k^{(in)}$(broken green line). \\ 
In all cases, the conditional mean is proportional to in-degrees (i.e., a power-law exponent of 1). 
(c)The CDFs of $x$ for Model-2 for the large time $t$ $(t=2^{15})$. 
Data shown are for the LSCC of the firm network (black solid line) and for the shuffled LSCC of the firm network. 
Sales of real firms scaled by the sum of sales are shown by the greed dash-dotted line.  
 For the firm networks, $x$ obeys the power law distribution with exponent about 1. The shuffled networks has with an exponent of 0.65, which corresponds to the exponent of the annealing approximation of the biased random walk model for the uncorrelated scale free network with the exponent 1.3. The CDF of $x$ agrees with the actual scaled sales for the large $s$, only for the firm network. \\
(d)The conditional mean of $x$ given in-degree $k^{(in)}$, denoted by $<x>_{k^{(in)}}$ for Model-1.
Data shown are for the LSCC of the firm network(black triangles), for the shuffled LSCC of the firm network(red diamonds), $x \propto {k^{(in)}}^{1.3}$(green dash-dotted line) and  $x \propto {k^{(in)}}^{2}$(blue dash-double-dotted line). }
\label{cdf_page}
\end{figure}

We apply the same analysis to Model-2.
Starting with the same initial condition as for Model-1, the CDF of $x$ converges quickly to a power law, as shown in figure \ref{cdf_page} (c) for both case of the LSCC of the real firm network and for the shuffled LSCC. 
The exponent of the steady state power law is about -1 for the LSCC of the firm network, which agrees with the empirical observation for sales, equation (\ref{cdf1}).
However, for the shuffled LSCC of the firm network, the exponent takes -0.65, which disagrees with the empirical observation equation (\ref{cdf1}). 
Note that the steady state of Model-2 is sensitive to the correlation of the network structure.
\par
figure. \ref{cdf_page} (d) shows the conditional mean of $x$ for given $k^{(in)}$ viewed as a function of $k^{(in)}$ , denoted by $<x>_{k^{(in)}}$.
For all cases, its behaviour approximately follows the power law:
\begin{equation}
<x>_{k^{(in)}} \propto \left\{ k^{(in)} \right\}^{\beta_{x|k}}, 
\end{equation}
with exponent $\beta_{x|k}$ satisfying equation (\ref{sisuu}). 
For the shuffled LSCC of the firm network, the exponent $\beta_{x|k}$, which takes a value about 2, is explained by an annealing approximation solution of the biased random walk model for an uncorrelated network \cite{bias}. However, this exponent disagrees with the empirical exponent given in equation (\ref{real_indegree_sales}). 
Conversely, for the LSCC of the firm network, the exponent $\beta_{x|k}$, which is equal to 1.3, agrees with empirical exponent given in equation (\ref{real_indegree_sales}).
These results indicate that the properties of the power-law exponent are connected to the space-correlation of the network. \par  

\section{Comparisons between simulations and observations} 
In this section, we consider the statistical properties of the entire firm network. 
For general networks, the steady states of the Model-1 and Model-2 do not always exist. In addition, because the scalar quantity defined for the nodes flows into nodes that do not have out-edges, most nodes in the network bulk have zero scalar after many time steps.
To obtain a non-trivial steady state we add the effects of injection and dissipation in the following forms: \\
Model-1 (PageRank model)
\begin{equation}
x^{}_{m}(t+1)=r\sum^{N}_{i=1}A_{im}\frac{1}{k_i^{(out)}}x^{}_{i}(t)+f , \label{model1b}
\end{equation}
Model-2 (Biased diffusion model)
\begin{equation}
x^{}_{m}(t+1)=r\sum^{N}_{i=1}A_{im}\frac{k_m^{(in)}}{\sum^{N}_{j=1}A_{ij}k_{j}^{(in)}}x^{}_{i}(t)+f \label{model2b}
\end{equation}
 where $0<1-r \leq 1$ is the dissipation factor and $f >0$ is the injection term, which is a constant and positive value. Note that, for $k_i^{(out)}=0$ in equation (\ref{model1b}) or $\sum^{N}_{j=1}A_{ij}k_{j}^{(in)}=0$ in equation (\ref{model2b}), we omit contributions of the i-th node.\par
In general, starting from any initial state, the time evolution given by $x(t+1)=rBx(t)+f$  converges to a unique steady state provided that the maximum eigen-value of $rB$ less than 1 \cite{linner}, where $x(t)$ is a state vector for time $t$ and $rB$ is a square matrix.
Denoting the maximum eigen-value of $rB$ as $\lambda$ and the corresponding eigen-vector as $y$ gives 
\begin{eqnarray}
|\lambda|\sum^{N}_{i=1}\left|y_i \right|&=&\sum^{N}_{i=1}|(rBy)_{i}|=r\sum^{N}_{i=1}\left|\sum^{N}_{j=1}B_{ij}y_j \right| \\
&\leq& r\sum^{N}_{i=1}\sum^{N}_{j=1} \left|B_{ji}y_i \right|=r\sum^{N}_{i=1}|y_i|\sum^{N}_{j=1}\left|B_{ji} \right| \label{koyuuti}
\end{eqnarray}
From this definition, 
we obtain the following for the Model-1
\begin{eqnarray}
\sum^{N}_{j=1}\left|B_{ji}\right|=\cases{
\sum^{N}_{j=1}A_{ij}\frac{1}{k_{i}^{(out)}}=\frac{k_{i}^{(out)}}{k_{i}^{(out)}}=1 & $(k_{i}^{(out)} \neq 0$) 
\\0 & ($k_{i}^{(out)} = 0$)  \\}
\end{eqnarray}
and for Model-2, we obtain
\begin{eqnarray}
\sum^{N}_{j=1}\left|B_{ji}\right|=  \nonumber \\
\cases{
\sum^{N}_{j=1}A_{ij}\frac{k_j^{(in)}}{\sum^{N}_{l=1}A_{il}k_{l}^{(in)}}=\frac{\sum^{N}_{j=1}A_{ij}k_{j}^{(in)}}{\sum^{N}_{l=1}A_{il}k_{l}^{(in)}}=1 & ($k_{i}^{(out)} \neq 0$) 
\\0 & ( $k_{i}^{(out)} = 0$)  \\ }
\end{eqnarray}
Thus, in both cases, 
\begin{equation}
r\sum^{N}_{j=1}|B_{ji}| < \sum^{N}_{j=1}|B_{ji}| \leq 1 \label{b1}
\end{equation}
Substituting equation (\ref{b1}) in equation (\ref{koyuuti}), we obtain
\begin{equation}
|\lambda|\sum^{N}_{i=1}\left|y_i \right| \leq \sum^{N}_{i=1}|y_i|\sum^{N}_{j=1}\left|rB_{ji} \right| <\sum^{N}_{i=1}|y_i|
\end{equation}
Thus, 
\begin{equation}
|\lambda| < 1
\end{equation}
Therefore, starting from any initial state, $x$ converges to a unique steady state. \par
\begin{figure}[t]
\begin{minipage}{0.3333\hsize}
\centering
\includegraphics[width=6cm]{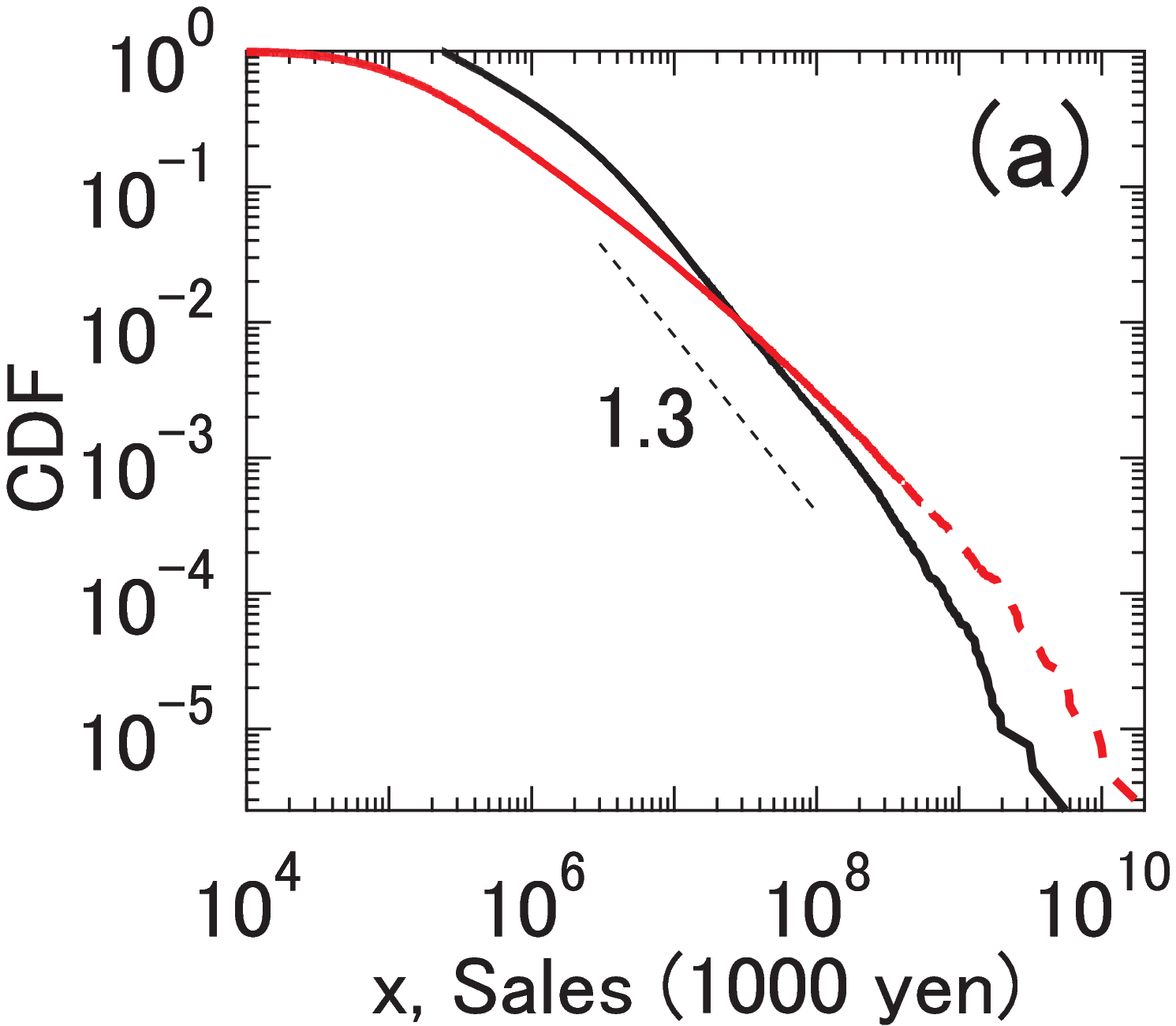}
\end{minipage}
\begin{minipage}{0.333333\hsize}
\centering
\includegraphics[width=6cm]{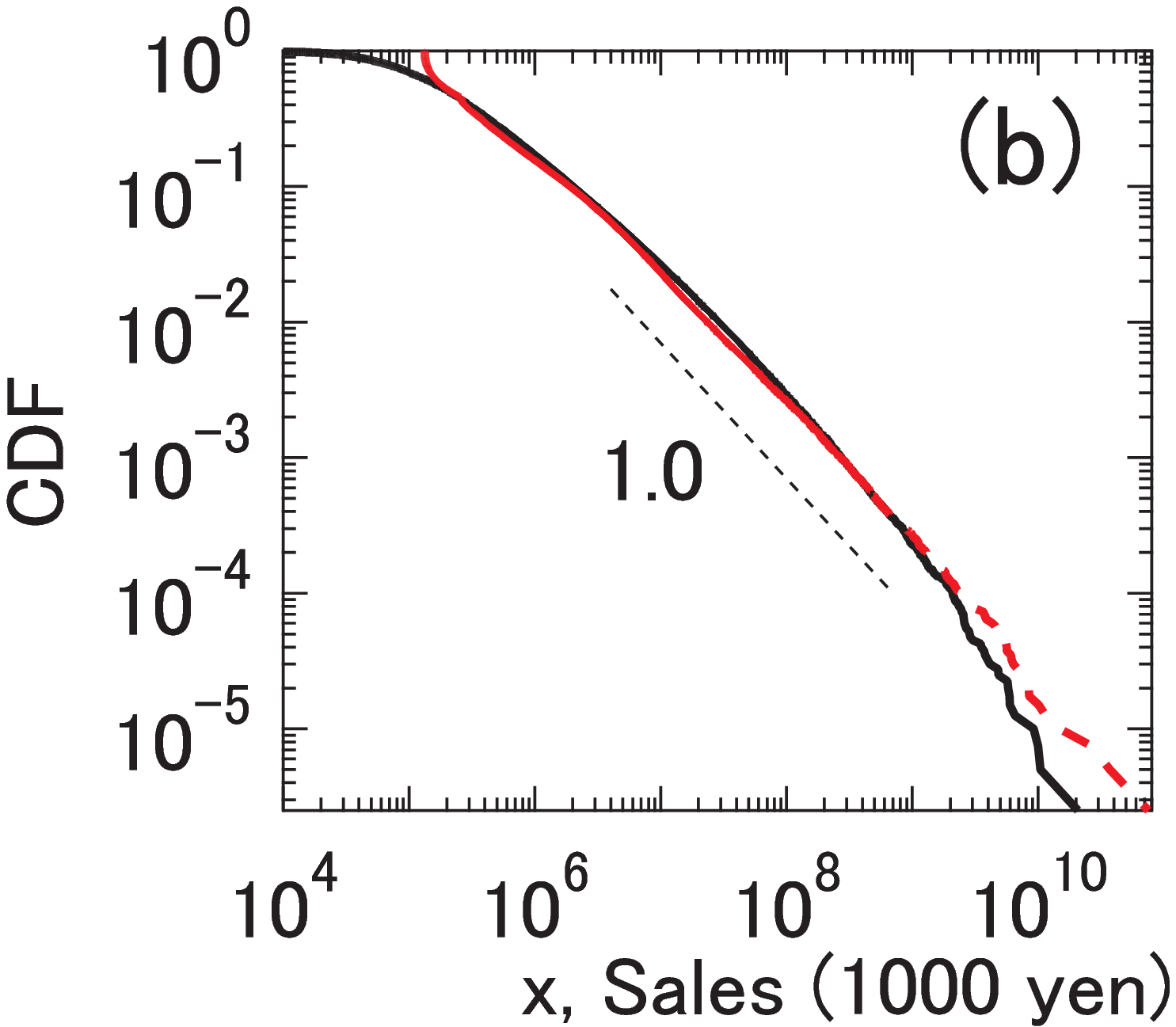}
\end{minipage}
\begin{minipage}{0.33333\hsize}
\centering
\includegraphics[width=6cm]{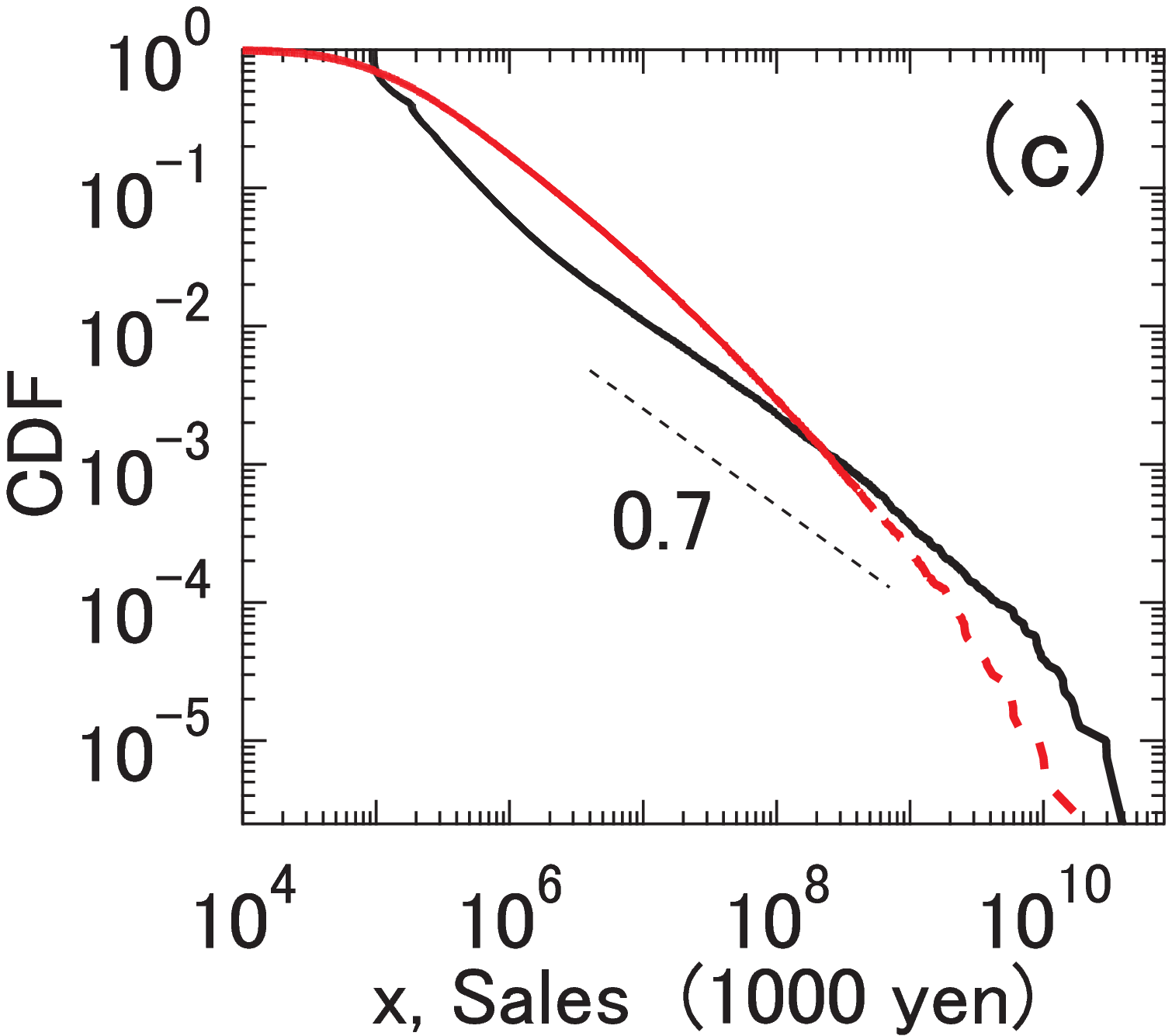}
\end{minipage}
\begin{minipage}{0.3333\hsize}
\centering
\includegraphics[width=6cm]{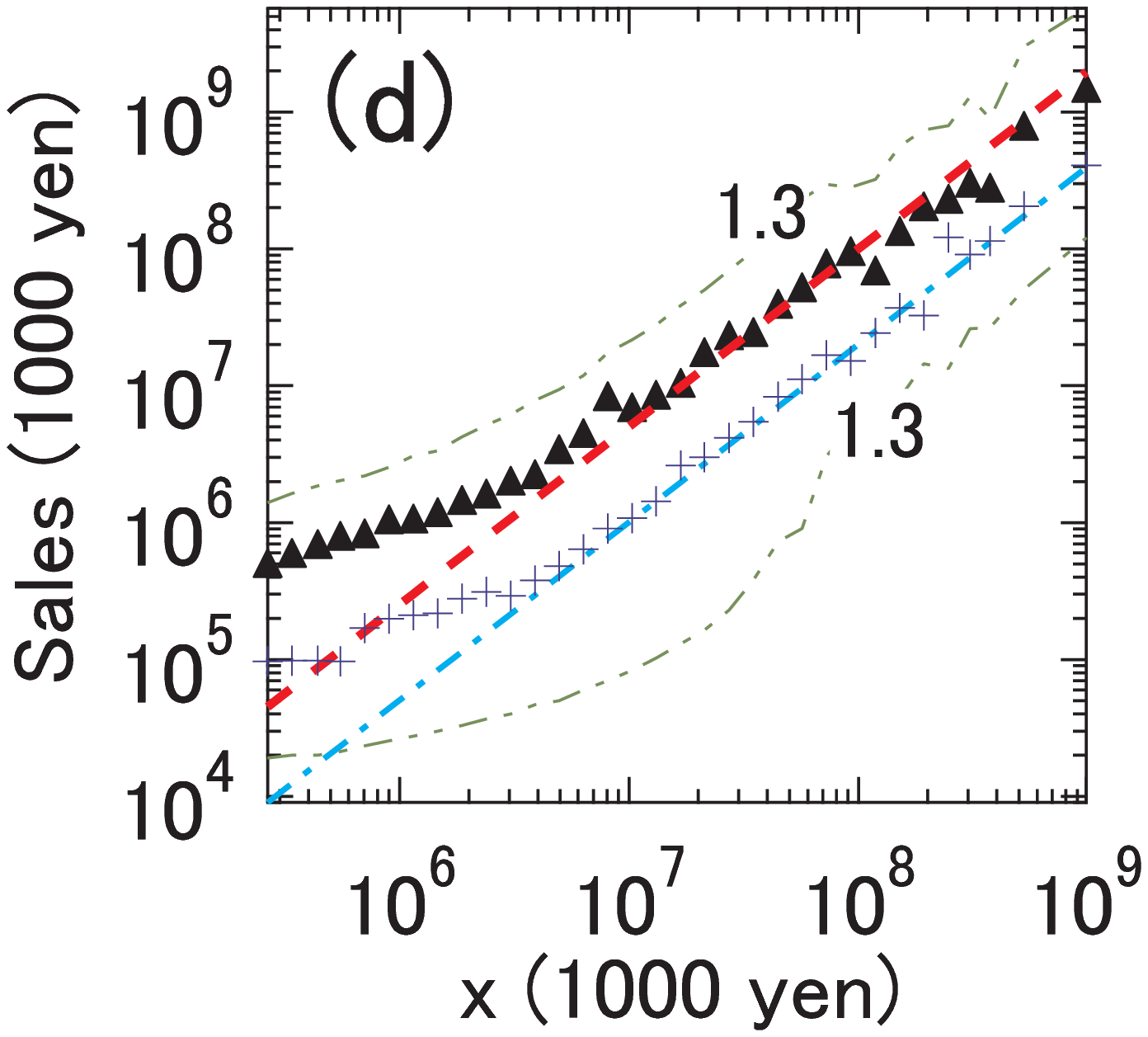}
\end{minipage}
\begin{minipage}{0.333333\hsize}
\centering
\includegraphics[width=6cm]{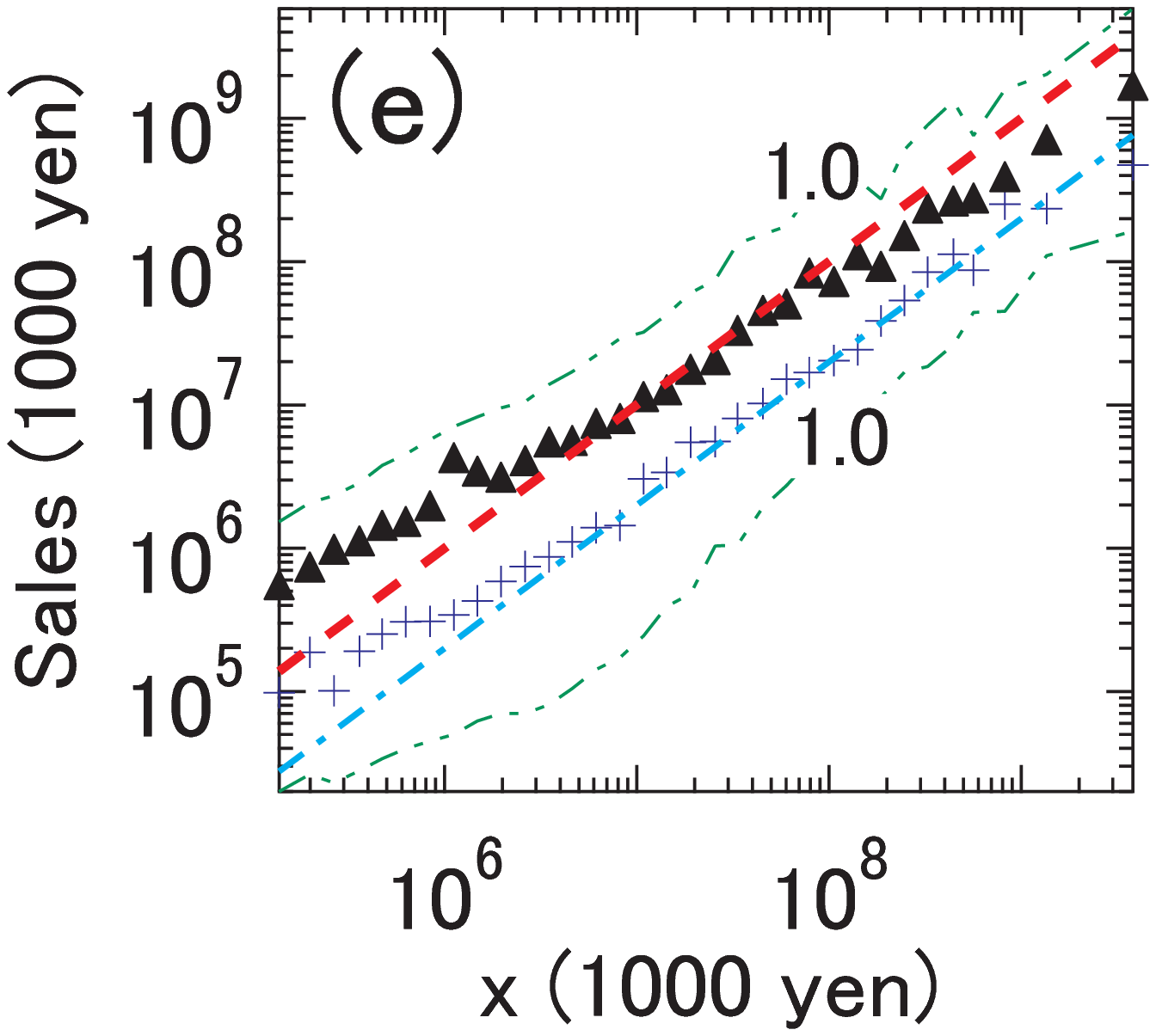}
\end{minipage}
\begin{minipage}{0.33333\hsize}
\centering
\includegraphics[width=6cm]{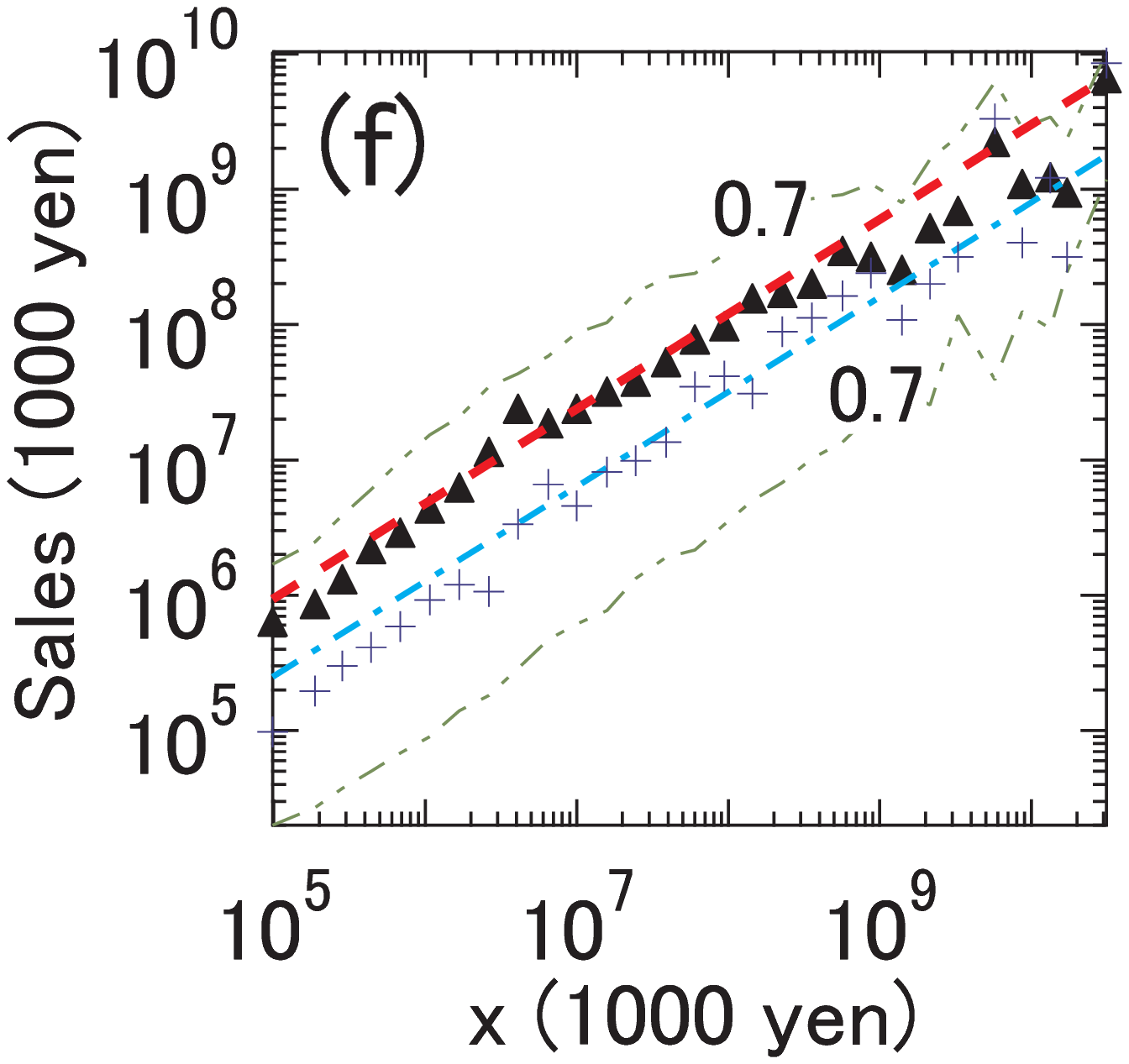}
\end{minipage}
 
\caption{
(a)-(c) Comparisons of the cumulative distribution between simulations and observations of actual sales. We use the value $r=0.95$ and $f=1.33 \cdot 10^5$ (1000 yen). 
Data shown are $x$ (solid black line) and actual sales (broken red line). (a)Model-1 for the firm network, (b)Model-2 for the firm network and (c)Model-2 for the shuffled network. 
For (a) and (c), the CDF of the sales $s$ disagrees with the CDF of simulation $x$; however, for (b), CDFs of the sales $s$ agrees with $x$. \\
(d)-(f) Correlation between $x$ and sales of actual firms $s$. Data shown are the conditional mean of sales $s$ given $x$, denoted by $<s>_{x}$ (black triangles), the conditional mode of sales given $x$ (blue crosses) and the conditional 5 percentile and the 95 percentile (green dashed-double-dotted line).  
(d)Model-1 for the firm network (broken red line: $s=0.004 \cdot x^{1.3}$, blue dash-dot line: $s=0.0008 \cdot x^{1.3}$), (e)Model-2 for the firm network (red broken line: $s=x$, blue dash-dotted line:$s=0.2 \cdot x$ ) and (f)Model-2 for the shuffled network (red broken line: $s=300 \cdot x^{0.7}$, blue dash-dotted line: $s=80 \cdot x^{0.7}$).
 }
 \label{cdf_hikaku}
\end{figure}

 
 In figure \ref{cdf_hikaku} we compare the CDFs between the simulations and observation. Figure \ref{cdf_hikaku}(a) shows the results for case of Model-1 for the firm network, figure \ref{cdf_hikaku}(b) shows a results for the case of Model-2 for the firm network and figure \ref{cdf_hikaku}(c) shows the results for Model-2 for the shuffled network. 
For these figures, we used $r=0.95$ and $f=1.33 \cdot 10^5$ (1000 yen). Under these conditions, $\sum_{i=1}^{N}s_i=\sum_{i=1}^{N}x_i(\infty)$ holds. 
From figure \ref{cdf_hikaku}(a) and (b), we see that the CDF of $x$ for Model-2 agrees well with the sales distribution observed for the firm network, whereas, the CDF of $x$ for Model-1 disagrees with the observed CDF.
In addtion, the result shown in figure \ref{cdf_hikaku}(c), implies that Zipf's law, which is observed in actual data, is strongly related to the degree-degree correlation mentioned in the preceding section. \par
Next, we compare the results for $x$ from our simulation with the sales of the real firms one by one.  
figure \ref{cdf_hikaku}(d), \ref{cdf_hikaku}(e) and \ref{cdf_hikaku}(f) show the actual firm sales on the vertical axis and the result for $x$ obtained from the above-mentioned network-flow on simulation in the horizontal axis.  Figure \ref{cdf_hikaku}(d) shows the result for Model-1 for the firm network, figure \ref{cdf_hikaku}(e) shows the same for Model-2 for the firm network and figure \ref{cdf_hikaku}(f) shows the result for Model-2 for the shuffled network.
 Figure \ref{cdf_hikaku}(e) show that for large $x$, in the case of Model-2, the conditional mean of actual sales $s$ given sales of the simulation $x$ as a function of $x$, denoted by $<s>_x$ is almost equal to $x$.
Meanwhile, figure \ref{cdf_hikaku}(d) shows that for Model-1 applied to the the firm network, $<s>_x$ is proportional to about the 1.3 power of $x$, whereas for Model-2 applied to the shuffled network, proportional to about the 0.7 power of $x$, as shown in figure \ref{cdf_hikaku}(f). In both cases, $<s>_x$ is not proportional to $x$. 
 This result suggests that Model-2(i.e., the model with injection and dissipation) applied to the firm network roughly reproduces the values of sales of actual firms for simulation sales $x$ larger than about $3 \cdot 10^6$ (1000 yen). \par

\section{Conclusion and discussion}
In this paper, we demonstrated that we can roughly estimate the sales of firms from the structure of the Japanese inter-firm trading network.
First  we found the simple linear local relationship between sales of a firm and the weighted sum of sales of its customers by analysing data from the Japanese inter-firm trading network and corresponding sales data. 
Next, we introduced a model (Model-2) that satisfies this local linear relationship between adjacent nodes.
 In this model, a firm (customer) distributes money to its out-edges (suppliers) proportionally to the in-degree of destinations.
By using this model to numerically simulate the real firm network, we confirmed that the steady flows  derived from the money-transport model reproduce the distribution of real firm sales and sales of individual firms on average. 
In addition, we also confirmed that the PageRank(Model-1), which corresponds to the equal distribution of money to out-edges, does not reproduce the distribution of sales.
Note that Model-2 corresponds to a biased random walk whose transition probabilities are
proportional to the in-degrees of destinations.
Therefore, based on our model, we argue that actual firm sales are proportional, on an average, to the existence probability (or the mean stay time) for the
steady state of a biased random walker on the firm network.
 \par
Applied to the firm network, Model-2 explains the difference between the exponent of the distribution of in-degrees and that of sales, and it also reproduces Zipf's law for the firm network. 
However, on application to the shuffled network, which is almost uncorrelated network with the same degree distribution as the real firm network, we also confirmed that sales derived by the model in the steady state does not obey Zipf's law. 
In other words, in our framework, we need not only a particular transport model but also a particular network structure to reproduce Zipf's law.
Thus, considering only the effect of the transport mentioned in this paper is sufficient to explain the universal Zipf's law. \par
For the  Model-2 simulation, the power-law exponent is influenced by the degree-degree correlation, whereas this dependence is not observed for Model-1 (PageRank model). 
Although many actual complex networks have the degree-degree correlation \cite{Breview}, few studies exist of biased random walk on correlated complex networks. A detail survey of this dependence will be reported in a future presentation. \par 
\ack
The authors are grateful to the Research Institute of Economy, Trade and industry (RIETI) for allowing us to use the TSR data. 
This work is partly supported by Grant-in-Aid for JSPS
Fellows Grant No. 219685 (H.W.) and Grant-in-Aid for
Scientiist Research No. 22656025 (M.T.) from JSPS.
The numerical calculations were carried out on TSUBAME at Global Scientific Information and Computing Center of Tokyo Institute of Technology.

\section*{References}

\end{document}